\documentclass[useAMS,usenatbib]{mn2e}
\usepackage{mathrsfs}
\usepackage{graphicx,amssym,color}
\def\lsim{\mathrel{\hbox{\rlap{\hbox{\lower4pt\hbox{$\sim$}}}\hbox{$<$}}}}
\def\gsim{\mathrel{\hbox{\rlap{\hbox{\lower4pt\hbox{$\sim$}}}\hbox{$>$}}}}
\newcommand{\bc}{\begin{center}}
\newcommand{\ec}{\end{center}}

\newcommand{\hMsun}          {\,h^{-1}\,{\rm M}_\odot}
\newcommand{\kms}            {\,\,{\rm km}\,\,{\rm s}^{-1}}

\title[Spatial Distribution of Galactic Satellites]
      {The Spatial Distribution of Galactic Satellites in the
      $\Lambda$CDM Cosmology}
\author[Jie ~Wang,  Carlos S. Frenk and Andrew P. Cooper]
       { Jie Wang$^1$ \thanks{Email: jie.wang@durham.ac.uk}, Carlos
         S. Frenk$^1$ and Andrew P. Cooper$^2$ \\
        $^1$Institute for computational cosmology, Department of Physics,
      Unidersity of Durham, Sourth Road, Durham, DH1 3LE, UK\\
       $^2$Max--Planck--Institut f\"ur Astrophysik,
        Karl--Schwarzschild--Str. 1, D-85748 Garching, Germany}
\begin{document}

\date{Accepted 2012 ???? ??.
      Received 2012 ???? ??;
      in original form 2007 ???? ??}

\pagerange{\pageref{firstpage}--\pageref{lastpage}}
\pubyear{2009}

\maketitle

\label{firstpage}

\begin{abstract} We investigate the spatial distribution of galactic
  satellites in high resolution simulations of structure formation in
  the $\Lambda$CDM model: the Aquarius dark matter simulations of
  individual halos and the Millennium II simulation of a large
  cosmological volume. To relate the simulations to observations of
  the Milky Way we use two alternative models to populate dark halos
  with ``visible'' galaxies: a semi-analytic model of galaxy formation
  and an abundance matching technique. We find that the radial density
  profile of massive satellites roughly follows that of the dark
  matter halo (unlike the distribution of dark matter subhalos).
  Furthermore, our two galaxy formation models give results consistent
  with the observed profile of the 11 classical satellites of the
  Milky Way. Our simulations predict that larger, fainter samples of
  satellites should still retain this profile at least up to samples
  of 100 satellites. The angular distribution of the classical
  satellites of the Milky Way is known to be highly anisotropic.
  Depending on the exact measure of flattening, 5--10 per cent of
  satellite systems in our simulations are as flat as the Milky Way's
  and this fraction does not change when we correct for possible
  obscuration of satellites by the Galactic disk. A moderate
  flattening of satellite systems is a general property of
  $\Lambda$CDM, best understood as the consequence of preferential
  accretion along filaments of the cosmic web.  Accretion of a single
  rich group of satellites can enhance the flattening due to such
  anisotropic accretion. We verify that a typical Milky Way-mass CDM
  halo does not acquire its 11 most massive satellites from a small
  number of rich groups. Single--group accretion becomes more likely
  for less massive satellites. Our model predictions should be
  testable with forthcoming studies of satellite systems in other
  galaxies and surveys of fainter satellites in the Milky Way.

\end{abstract}

\begin{keywords}

\end{keywords}

%%%%%%%%%%%%%%%%%%%%%%%%%%%%%%%%%%%%%%%%%%%%%%%%%%%%%%%%%%%%%%%%%%
\section{Introduction}

The satellites of the Milky Way offer a number of critical tests of
the current cosmological paradigm, the $\Lambda$CDM model. Their
abundance and internal structure are sensitive to the nature of the
dark matter and thus to the high frequency end of the linear power
spectrum of density fluctuations. Their spatial distribution is
sensitive to the gravitational evolution of dark matter on galactic
and supergalactic scales.

High resolution simulations of the formation of galactic cold dark
matter halos have revealed that a large number of substructures
survive to the present day, accounting for about 10\% of the total
halo mass \citep[][and references therein]{diemand07,springel08}.
Since only about two dozen satellites are known to orbit in the halo of
the Milky Way, this property is frequently deemed to pose a
``satellite problem'' for CDM-based cosmologies. In fact, it was shown
about a decade ago that basic processes that are unavoidable during
galaxy formation, such as supernova feedback and early Hydrogen
reionization, readily explain why a visible satellite galaxy can form
only in a tiny fraction of the surviving subhalos
\citep{bullock00,benson2002,somerville2002}. This result has been
confirmed repeatedly in recent years using semi-analytic models
\citep{cooper10,guo11,li10,maccio10,font11}, cosmological gasdynamical
simulations \citep{okamoto_frenk10,parry12,wadepuhl11} and simplified
semi-empirical models \citep{kravtsov04,koposov09,busha10,munoz09}.

A different kind of theoretical challenge is posed by the spatial distribution
of the 11 classical satellites of the Milky Way.  \cite{lyndenbell76},
\cite{kunkel79} and \cite{lyndenbell82} noted that these satellites lie very
close to a great circle on the sky that is approximately perpendicular to the
Galactic Plane. \cite{kroupa05} deemed such a highly flattened structure to be
extremely unlikely in a CDM cosmology but, using N-body simulations of halo
formation in the $\Lambda$CDM model, \cite{kang05}, \cite{libeskind05},
\cite{zentner05} and \cite{libeskind09} showed explicitly that this presumption
is incorrect. Such flattened satellite distributions were dubbed ``great
pancakes'' by \cite{libeskind05} who ascribed them to highly anisotropic
accretion of proto-subhalos along the filaments of the cosmic web.
Correlated accretion along filaments was also identified as the cause for
the polar alignment of satellite disks found by \cite{deason11} in 20 percent
of satellite systems (with more than 10 bright members each) in the {\small
``GIMIC''} N-body/gasdynamic simulations \citep{crain09}. Although all of
these studies found that flattening of satellite systems is common in
$\Lambda$CDM, they also found that the high degree of flattening in the Milky
Way system is atypical.

\cite{li08} showed that the flattening effects of anisotropic
accretion are greatly enhanced in cases where infalling dark matter
subhalos belong to a `group' sharing similar infall times and orbital
angular momentum orientations (the members of such groups do not have
to be bound in a single DM halo before infall). In a Mikly Way-mass
system, they found that samples of $\sim10$ subhalos drawn from one or
two such groups readily produced configurations as flat as that of the
classical Milky Way satellites. Extrapolating this result, they
suggested that the highly correlated infall of a rich group of
satellites was a {\em possible} explanation for the abnormal degree of
flattening seen in the Milky Way.  However, they could not say whether
this was a {\em probable} explanation since they did not calculate the
likelihood of all the 11 {\em brightest} satellites being members of
only one or two such groups.

An important limitation of the simulations that have been analysed so far to
study the spatial distribution of satellites is their relatively low
resolution. Low resolution can give rise to excessive tidal disruption and to
the artificial merging of some subhalos, potentially obscuring the true spatial
distribution. In this paper, we use the suite of six simulations of individual
Galactic halos from the Aquarius project, which are amongst the highest
resolution CDM simulations carried out to date \citep{springel08}, as well as a
sample of similar halos from the Millennium-II simulation \citep{boylan09}.
Millennium-II has lower resolution than Aquarius but follows halo formation in
a cosmological volume (a cube of side 100 $h^{-1}$Mpc, where $h$ denotes the
Hubble constant in units of 100 ${\rm km s}^{-1}{\rm Mpc}^{-1}$). With the
Aquarius simulations we are able to study satellites down to very small masses
in six galactic halos and with the Millennium-II we are able to study the
massive satellites of a large sample of galactic halos. Since the Aquarius
halos are zoom-resimulations of regions in the Millennium-II volume, we
are able also to study the effects of numerical resolution.

We rank satellites in our simulations by stellar mass using two different
techniques: semi-analytic modelling and a simple assignment of the brightest
satellites to the most massive proto-subhalos. As we show, the two approaches
pick out similar subsets of subhalos as satellite hosts. With these samples, we
revisit the radial distribution of satellites and the great pancake and, for
the first time, we investigate how these properties depend on the number of satellites considered (in samples ranked by stellar mass). This
allows us to make predictions for forthcoming surveys such as Pan-STARRS
\citep{kaiser10}, which may discover a large population of very faint
satellites in the Milky Way and M31, and the ongoing Pandas survey of M31
\citep{mcconnachie09, martin09}. We also extend the work of \citet{li08}
by using Aquarius to investigate the multiplicity function of groups of massive
satellites.

The paper is organized as follows. In Section~\ref{sec:model}, we discuss the
suite of simulations and galaxy formation models that we use. In
Section~\ref{sec:results}, we investigate the radial distribution of
satellites, the prevalence of great pancakes, and groups of massive satellites.
Our conclusions and discussion are presented in Section~\ref{sec:conclusion}.

\begin{figure}
\bc
\hspace{1.cm}
\resizebox{10cm}{!}{\includegraphics{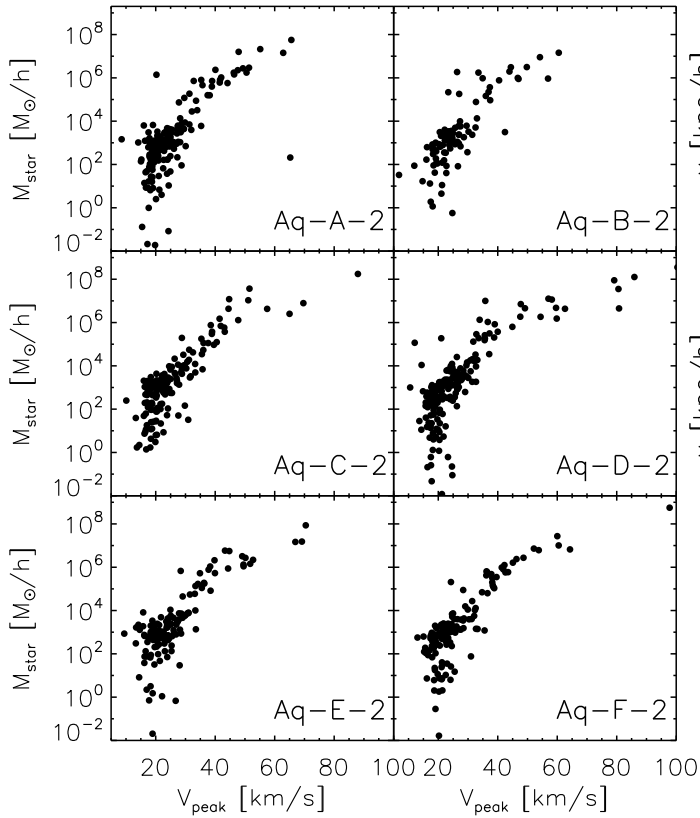}}\\%

\caption{The relationship between the stellar mass assigned by the
\citet{cooper10} semi-analytic model to subhalos in the six Aquarius
simulations and the maximum value of the rotation curve attained by each
subhalo over its entire past history. (Satellites with unresolved dark
matter haloes are excluded.) The tight correlation for the most massive
satellites motivates a simple $V_{\rm peak}$ model for ranking satellites by
stellar mass.}

\label{fig:Fig1}
\ec
\end{figure}

\begin{figure}
\bc
\hspace{1.cm}%
\resizebox{10cm}{!}{\includegraphics{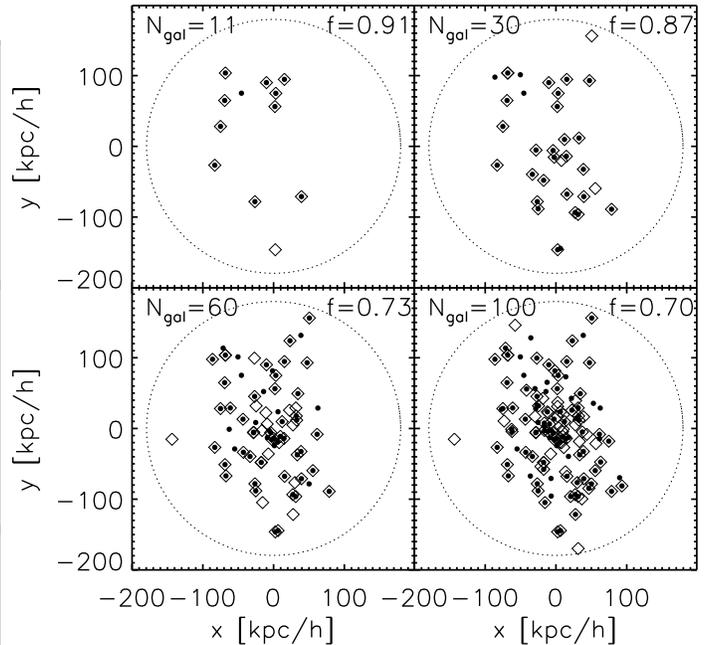}}\\%

\caption{The projected positions of satellites identified using the
\citet{cooper10} semi-analytic model (diamonds) and the $V_{\rm peak}$ method
(filled circles) in the Aq-A2 halo. From top left to bottom right, positions are shown for the
top 11, 30, 60 and 100 satellites ranked by stellar mass in each model.
The overlap ratio between the two samples is given in the legend. It decreases
from $f=0.91$ for the top 11 to $f=0.7$ for the top 100 satellites.}

\label{Fig2}
\ec
\end{figure}

\section{Satellite identification}
\label{sec:model}

We begin by providing a brief introduction to the Aquarius and
Millennium-II simulations and then describe the two alternative
techniques we have employed to identify satellites in them.

\subsection{Aquarius and Millennium II}

The Aquarius Project \citep{springel08} consists of a suite of very
high resolution N-body simulations of six dark matter halos of mass
similar to that expected for the halo of the Milky Way.  The
simulations assume the $\Lambda$CDM cosmology, with parameters
consistent with the {\small WMAP}~1 data
\citep{spergel03}: matter density parameter, $\Omega_{\rm M}=0.25$;
cosmological constant term, $\Omega_{\Lambda}=0.75$; power spectrum
normalization, $\sigma_8=0.9$; spectral slope, $n_s=1$; and Hubble
parameter, $h=0.73$. These values are inconsistent with the {\small
WMAP}~7 data at about the 2$\sigma$ level, but this discrepancy is of
no consequence for the analysis of this paper \citep[see][]{wang12}. 

The six Aquarius halos are labelled ``Aq-A'' through ``Aq-F''. Each
was resimulated at different resolution in order to assess
numerical convergence. A suffix 1 to 5 identifies the
resolution level, with level 1 denoting the highest resolution.
In this study, we analyse primarily the
level 2 simulations.
%, but we also use lower resolution versions to test for numerical
%convergence.
For the level 2 simulations, the
particle mass is $m_{p}\simeq 1 \times 10^4\hMsun$ and the softening
length is $ \epsilon= 48 h^{-1} {\rm pc}$. At $z=0$, the six halos
have a ``virial'' mass, $M_{200}\sim 1$-$2 \times
10^{12}\hMsun$, where $M_{200}$ is the mass contained within
$r_{200}$, the radius of a sphere of mean density 200 times the
critical density for closure. The circular velocity curves of the
halos peak at $V_{\rm max} =220 \pm40 \kms$.  Although the
Aquarius halos have similar final masses, they have varied
formation histories \citep{wang11}. For further details of the
Aquarius Project, we refer the reader to \citet{springel08} and
\citet{navarro10}.

At every snapshot in the simulations we find nonlinear structures
using the friends-of-friends (FOF) algorithm of \citet{davis85}, with
a linking length of 0.2 times the mean interparticle separation and
$32$ particles as the minimum number of particles per group.  We also
identify bound substructures within each FOF halo using
the {\small SUBFIND} algorithm of \citet{springel05}. Merger trees for
subhalos are constructed as described in \citet{springel08}.

The Millennium-II run (MRII) is a cosmological simulation in which
$2160^3$ particles were followed in a cubic box of side length $L_{\rm
box} = 100~{\rm Mpc}~h^{-1}$. This volume is 125 times smaller than that of
the Millennium simulation \citep{springel05}, and the mass resolution is correspondingly
125 times better: each particle has mass $6.88 \times 10^6
h^{-1}M_{\odot}$. The cosmological model is the same as that assumed
for the Aquarius Project. For further details, we refer the reader to
\citet{boylan09}.

The Aquarius halos are resimulations of regions selected from a low resolution
version of the MRII with identical large-scale density perturbations and
phases in their initial conditions. We can therefore find the counterparts of
each Aquarius halo in the MRII and carry out resolution tests
\citep[see][]{boylan09}. The resolution of the MRII is similar to that of the
simulations analysed by \cite{kang05} and \cite{zentner05}, but it is a factor
of $\sim30$ worse than the simulation by \cite{libeskind05} and a factor of
$\sim20$ better than that by \cite{libeskind09}.

\subsection{Galaxy formation models}

There are two techniques for modelling satellite galaxies in detail
using simulations: direct hydrodynamic simulations and semi-analytic
modelling applied to a high-resolution N-body simulation of the dark
matter. The two techniques are similar, differing primarily in the
simplified, spherically symmetric treatment of gasdynamics in the
semi-analytic method. In this study, we are interested in very faint
satellites. Even the highest resolution gasdynamic simulations of
satellites performed to date \citep{okamoto_frenk10,
wadepuhl11} can resolve only the dozen or so brightest galaxies pf a
Milky Way system and so
they are not suitable for our purposes. Semi-analytic models, on the
other hand, are restricted only by the resolution of the associated
N-body simulation so we resort instead to the semi-analytic models
applied to the Aquarius simulations by \cite{cooper10} and to the MRII
by \cite{guo11}. We refer the reader to these papers for details of
the implementation. An added advantage of the semi-analytic models is
that, unlike the direct gasdynamic simulations, these models are known
to agree with a whole range of other properties of the galaxy
population at various epochs.

In practice, despite the complexity of the many physical processes they
incorporate, both gasdynamic and semi-analytic simulations predict an
approximately monotonic relation between the stellar mass of a galaxy and the
total mass of the halo in which it forms. As we are only interested in
the stellar mass rank of satellites, it is instructive to consider a much
simpler model for satellites in which the stellar mass (hence
luminosity) of each satellite is assumed to be proportional to the highest
value of the maximum of the circular velocity curve ($V_{\rm max}$) attained by
the halo in which the satellite forms throughout its entire history. We denote
this by $V_{\rm peak}$; it corresponds roughly to the value of $V_{\rm max}$
just prior to the halo being accreted into the main halo and becoming a
subhalo. We refer to this as the ``$V_{\rm peak}$ model''. By comparing
our semi-analytic results with this simpler but fundamentally similar model, we
can check if our conclusions depend on the details of the semi-analytic
treatment, which introduces a scatter in the relationship between stellar mass
and (maximum) halo mass.

The correspondence between $V_{\rm peak}$ and $M_{\rm star}$, the
stellar mass of  the semi-analytic model of \cite{cooper10}, is illustrated in
Fig.~\ref{fig:Fig1}. The expected monotonic relation is apparent. However,
there is considerable scatter in the relation for subhalos with $V_{\rm
peak}\lsim 25 {\rm km s}^{-1}$. At higher masses, in the range of the 11
classical satellites, the relation flattens and the scatter is greatly reduced.
In this regime, over 90\% of the satellites with the highest $V_{\rm peak}$
values also have the highest $M_{\rm star}$ according to the semi-analytic
model. Fig.~\ref{fig:Fig1} shows that our simple $V_{\rm peak}$ model will give
a ranking of subhalos by stellar mass very similar to that of the
semi-analytic model, in most cases.

The internal structure of halos that are close to the resolution limit of an
N-body simulation is, of course, poorly modelled. This can lead to the
artificial disruption of these halos when they are accreted into larger halos.
The baryonic properties of their associated galaxies can still be followed in
semi-analytic models even after the subhalos fall below the resolution of the
simulation, but as their orbits cannot be tracked within the N-body simulation
their spatial distribution is uncertain and model-dependent.    
The resolution of the
Aquarius simulations is so high that we need not worry about galaxies with
unresolved haloes \citep[see discussion of this point in][]{font11}. However,
at the lower resolution of the MRII, galaxies in unresolved haloes do need to
be taken into account, as discussed by \citet{guo11}. This conclusion is
supported by our investigation of the radial distribution of satellites in MRII
in the following section. We demonstrate that convergence with Aquarius is only
possible if we include galaxies without resolved subhaloes in the Guo
et al. model. In
addition, because the circular velocity scale corresponding to the
$V_{\mathrm{peak}}$ of the most massive satellites ($30$--$60\, \mathrm{km \,
s^{-1}}$) is close to the resolution limit of MRII, we can only make reliable
comparisons between the semi-analytic and the $V_{\rm peak}$ models in
the Aquarius simulations, and not in the MRII.

In Fig.~\ref{Fig2} we compare the projected positions of the top 11, 30, 60,
and 100 satellites ranked by stellar mass in Aq-A-2 according to the
semi-analytic model (diamonds) and the $V_{\rm peak}$ method (full circles). As
we can see in the top-left panel, 10 out of the top 11 satellites are the same
in both models, an overlap ratio, $f=10/11=0.91$.  As we consider smaller and
more numerous satellites, $f$ decreases, but even for the top 100 satellites,
$f$ is still as high as 70 percent.

\begin{figure}
\bc
\hspace{-1.4cm}
\resizebox{9cm}{!}{\includegraphics{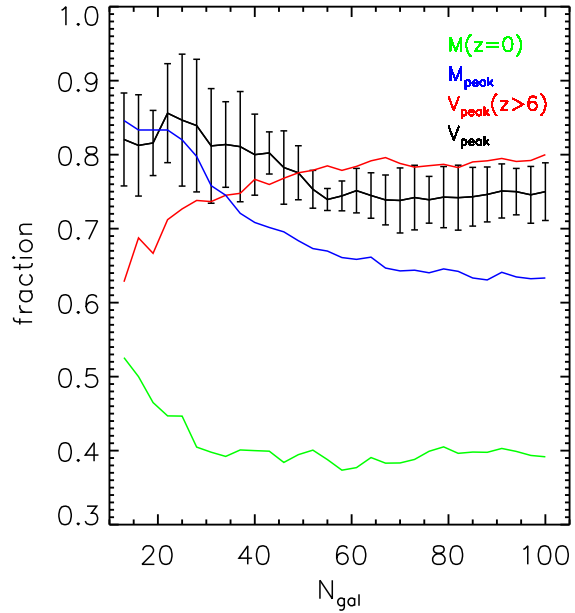}}\\%

\caption{The fraction of satellites ranked according to different criteria that
overlap with the top $N_{gal}$ satellites ranked according to the stellar mass
assigned to them by the \citet{cooper10} semi-analytic model. The data
correspond to the mean value for all six Aquarius halos. The black line shows
the overlap with our standard $V_{\rm peak}$ model and the 1$\sigma$ error bar
obtained from the 6 Aquarius simulations. The green line shows the result of
ranking satellites according to their present mass and the blue line according
to the maximum mass ever attained by the halo in which they reside today. The
red line is similar to the $V_{\rm peak}$ model, except that in this case
$V_{\rm peak}$ is defined as the maximum value attained by $V_{\rm max}$ before
$z=6$, approximately the redshift of reionization.}

\label{fig:Fig3}
\ec
\end{figure}

In Fig.~\ref{fig:Fig3} we compare the overlap fractions between
the top $n_{gal}$ satellites ranked according to the stellar mass
assigned by the \cite{cooper10}  model and various simplified but
plausible models for identifying satellites with subhalos, including
our standard $V_{\rm peak}$ model (black line with error bars).  The
overlap fraction for the $V_{\rm peak}$ model tends to a constant,
$f\simeq 0.75$, for $n_{gal}\gsim 60$. The green line corresponds to
the case in which satellites are identified with the most massive
subhalos at the present day \citep{stoehr02}. Clearly, this method
fails to place satellites in similar subhalos to those picked out by
our two standard models. The reason for this is simply that the
present mass of a subhalo is significantly affected by tidal stripping
and thus is only weakly correlated with the mass of the halo prior to
accretion, during which time most of the satellite stars form. 

Identifying satellites instead with subhalos ranked by the maximum mass they
ever attain (blue line) gives a better match to the standard models since this
mass is better correlated with the satellite's stellar mass or luminosity. Even
in this case, though, the overlap falls to just over 60 percent for large
satellite populations. The red line shows what happens if we identify
satellites according to the value of $V_{\rm peak}$ attained prior to $z=6$,
roughly the redshift of reionization. This model gives similar results to our
standard $V_{\rm peak}$ model, except for the most massive satellites,
reflecting, in part, the fact that reionization plays a relatively minor role
in setting the stellar mass of these galaxies \cite[see
e.g.][]{okamoto_frenk10,font11}. We conclude that our choice of
$V_{\mathrm{peak}}$ is the most appropriate of these alternatives and
in mos of the
remainder of this paper we focus on comparisons between satellites ranked by
this property and by the stellar mass in our semi-analytic models.

\begin{figure}
\bc
\hspace{-1.4cm}
\resizebox{9cm}{!}{\includegraphics{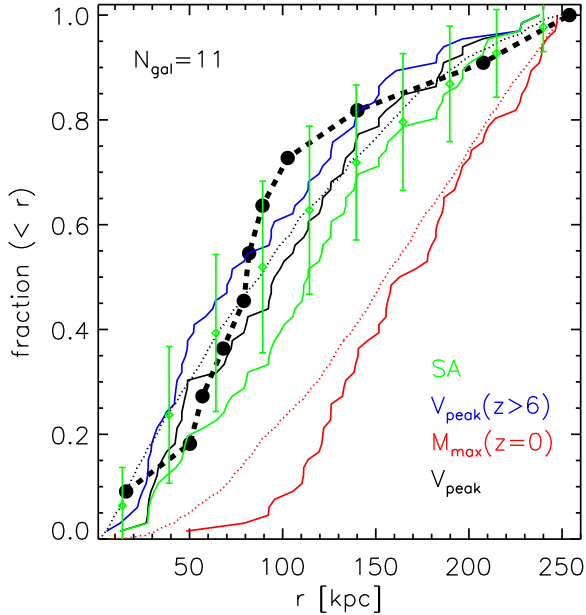}}\\%
\caption{The radial profiles of the top 11 satellites in different
  models. Results have been averaged over the 6 Aquarius halos. The
  profile of the 11 classical satellites of the Milky Way is shown by
  the filled circles joined with a thick dashed black curve. The
  radial profile of the dark matter is shown by the black dotted curve
  and the radial profile of all resolved subhalos in the six Aquarius
  halos by the red dotted curve. The green, blue, red and black lines
  show the distribution of satellites in the Aquarius simulations,
  identified using different models, as indicated in the legend. The
  green diamonds with error bars correspond to satellites in the
  semi-analytic model applied to the MRII, including satellites
  without resolved dark matter haloes.}
\label{fig:Fig4}
\ec
\end{figure}

\begin{figure}
\bc
\hspace{-1.4cm}
\resizebox{9cm}{!}{\includegraphics{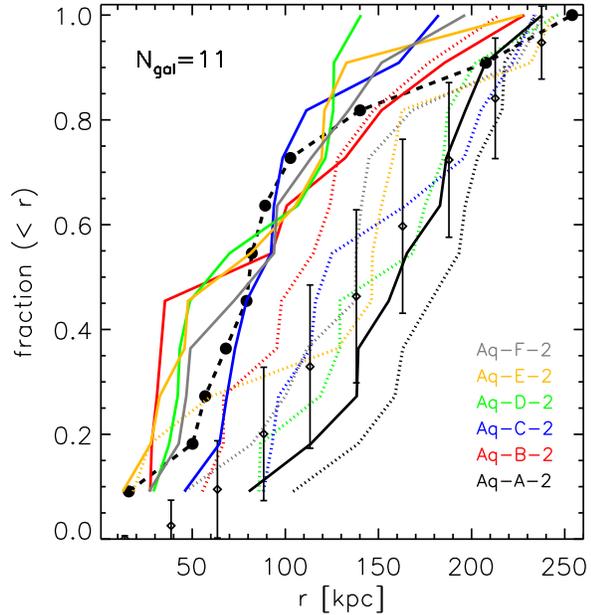}}\\%
\caption{The radial profile of the 11 brightest satellites according
  to the $V_{\rm peak}$ model in each of the six Aquarius halos (solid
  lines) and in their counterparts in the MRII (dotted lines). The
  diamonds and large error bars correspond to the average profile and
  1$\sigma$ dispersion of the top 11 satellites identified by the
  $V_{\rm peak}$ model in 1686 MRII halos of mass $1\times 10^{12}
  M_{\odot} < M_{200} < 2 \times 10^{12} M_{\odot} $. The profile of
  the 11 classical Milky Way satellites is shown by the black circles
  joined with a dashed black curve.}
\label{fig:Fig5}
\ec
\end{figure}

\begin{figure}
\bc
\hspace{1.4cm}
\resizebox{10cm}{!}{\includegraphics{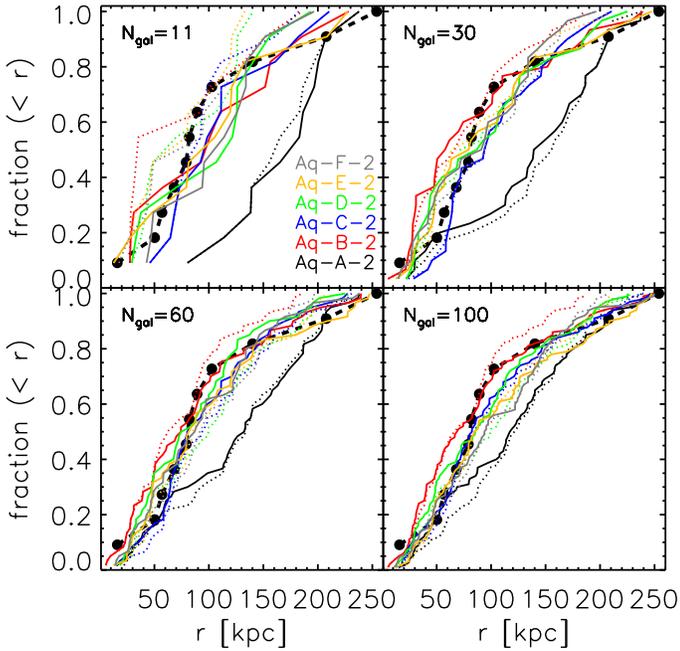}}\\%
\caption{The cumulative radial profiles of the top $N_{\rm gal}$
  satellites in each of the Aquarius halos in the semi-analytic (solid
  lines) and $V_{\rm peak}$ (dotted lines) models.  The profile of the
  11 classical satellites of the Milky Way is shown in all panels by
  the solid dots joined by a thick dashed black curve.}
\label{fig:Fig6}
\ec
\end{figure}

\section{Results}
\label{sec:results}
 
In this section we consider the spatial distribution of satellite
galaxies predicted in both our semi-analytic and $V_{\rm peak}$ models
and compare them with data for satellites in the Milky Way.

\subsection{Radial profiles}

The radial distribution of the 11 classical satellites of the Milky Way is
shown by the filled circles in Fig.~\ref{fig:Fig4}. The Galactocentric distance
used here is taken from \cite{mateo98}.  It is close to the radial distribution
of the dark matter in the Aquarius halos, shown in the figure by the black
dotted line, a similarity already noted by \cite{kang05}, \cite{libeskind05}
and \cite{font11}. To compare the data to our models, we select the top 11
satellites in each model that lie within a distance of 250~kpc from the centre
of the halo, roughly the distance of Leo I, the most distant of the
observed Milky Way eleven.

Results for the semi-analytic, $V_{\rm peak}$, $V_{\rm peak}(z>6)$ and
$M_{\rm max} (z=0)$ models are shown by the green, black, blue and red
solid lines respectively. All, except the last, are roughly consistent
with the data within the large uncertainties due to small number
statistics \citep[see also][]{cooper10,font11}. By contrast,
identifying satellites with the most massive present day subhalos (red
line) leads to a much more extended distribution which, as first noted
by \cite{libeskind05}, closely follows the radial profile of the
entire halo population. This is consistent with the finding by
\cite{springel08} that the radial profile of subhalos depends very
weakly on subhalo mass. The other, more realistic, models place bright
satellites in subhalos whose radial profiles are strongly biased
relative to that of the subhalo population as a whole and which happen
to lie close to the dark matter radial profile. We also show results
for Milky Way galaxies from the semi-analytic model applied to the
MRII, taking care to include satellites which survive even though
their dark matter halos have been tidally stripped. The distrobution
is close to the observational data. However, had we not included
satellites without resolved dark matter halos, the profile would lie
close to that of all subhaloes (red dotted).

The model lines in Fig.~\ref{fig:Fig4} are smoother than the Milky Way data
because we averaged the distributions of the 11 top-ranked satellites from
each of the 6 Aquarius halos. There is, in fact, considerable halo to halo
scatter, as may be seen in Fig.~\ref{fig:Fig5}, which shows the radial profiles
of the 11 brightest satellites predicted by the semi-analytic model in each
halo. Five of these lie very close to the Milky Way data but in Aq-A, the
radius enclosing half the satellites is over twice as large as in the
other 5 Aquarius halos. 

Fig.~\ref{fig:Fig5} also highlights the importance of high resolution
simulations for this kind of study. Since the Aquarius halos were selected from
the MRII, we can identify their counterparts in that simulation, as described
by \cite{boylan10}. The radial profiles of the top 11 satellites in these
counterparts, as identified in the $V_{\rm peak}$ model, are shown by the
dotted lines in the figure, with colours corresponding to those of the Aquarius
halos.  Satellites in the MRII counterparts are clearly much more extended than
the Aquarius subhalos. This is typical of the satellite distributions in Milky
Way-mass halos in the MRII, as illustrated by the diamonds with error bars,
obtained by averaging the distributions of the 11 brightest satellites in the
1686 MRII halos of mass $1 \times 10^{12} \leq M_{\rm 200} \leq 2 \times
10^{12}$. The MRII has about 1000 times
lower resolution than the level 2 Aquarius simulations. As a result, subhalo
orbits are not followed as accurately and their disruption timescale is
artificially reduced. 

The semi-analytic model of \citet{guo11} allows galaxies to survive
after their host subhalo falls below the resolution limit. A position
is assigned to these galaxies by tracking the most-bound particle of
the host subhalo from the time it was last resolved.  This position is
unlikely to be a very accurate estimate of the true orbit of the
satellite; for example, Guo et al. make an analytic correction to this
single-particle orbit when calculating the tidal disruption of the
satellite, as the orbits of individual particles do not decay through
dynamical friction in an N-body simulation.  Nevertheless, when we
include satellites with unresolved haloes in our top 11 sample, as was
done in Fig.~\ref{fig:Fig4} (but {\em not} in Fig.~\ref{fig:Fig5}),
and use their single-particle position estimates, the MRII radial
profiles for the Aquarius galaxies are in reasonable agreement with
their higher resolution counterparts.

As we expect that larger and more complete samples of fainter satellites
will soon be available for the Milky Way and other galaxies, it is interesting
to know how the radial profiles of satellites are predicted to vary with the
luminosity cut of the sample. This is a question that we can address with the
Aquarius simulations and we do this in Fig.~\ref{fig:Fig6}, where we show the
radial profiles of the top 11, 30, 60 and 100 galaxies predicted in the
semi-analytic (solid curves) and $V_{\rm peak}$ (dotted curves) models. The
profiles depend only weakly on the size of the sample within the mass
range we can examine: they are all similar to that of the 11 classical
satellites of the Milky Way. The exception is Aq-A whose profiles become
increasingly less discrepant with the other 5 Aquarius cases as the satellite
sample size increases (although it is still the least concentrated
even when 100 satellites are considered). By most other measures Aq-A is
representative of haloes of the same mass (\citealt{boylan10}), suggesting that
the top 11 satellites may often exhibit `unusual' configurations in otherwise
`typical' haloes.

\begin{figure}
\bc
\hspace{-1.4cm}
\resizebox{9cm}{!}{\includegraphics{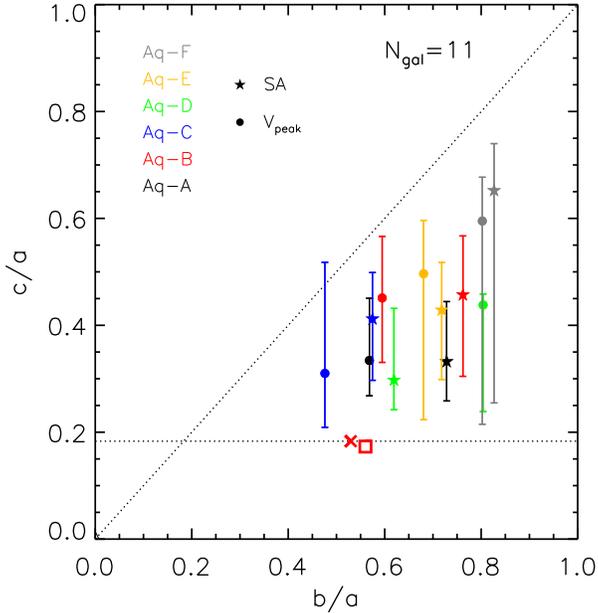}}\\%
\caption{The axis ratios describing the shape of satellite systems in
  the Aquarius simulations and in the Milky Way. $c/a$ is the minor to
  major axis ratio and $b/a$ the intermediate to major axis ratio. Results
  for the top 11 satellites in the semi-analytic and $V_{\rm peak}$
  models are shown by asterisks and filled circles respectively, with
  error bars spanning the range obtained for a large sample of
  satellites taking into account systematic effects introduced by the
  zone of avoidance (see text for details). The axis ratios for the 11
  classical satellites of the Milky Way are shown with a red cross
  with the value of $c/a$ highlighted by the horizontal dotted
  line. The average axis ratios of the 6 Aquarius dark matter halo
  (within $r_{200}$) are shown as the large open triangle towards
  the top-right corner. The red square shows the effect of including
  Canes Ventici in the observational sample.}
\label{fig:Fig7} 
\ec
\end{figure}

\begin{figure}
\bc
\hspace{-1.4cm}
\resizebox{9cm}{!}{\includegraphics{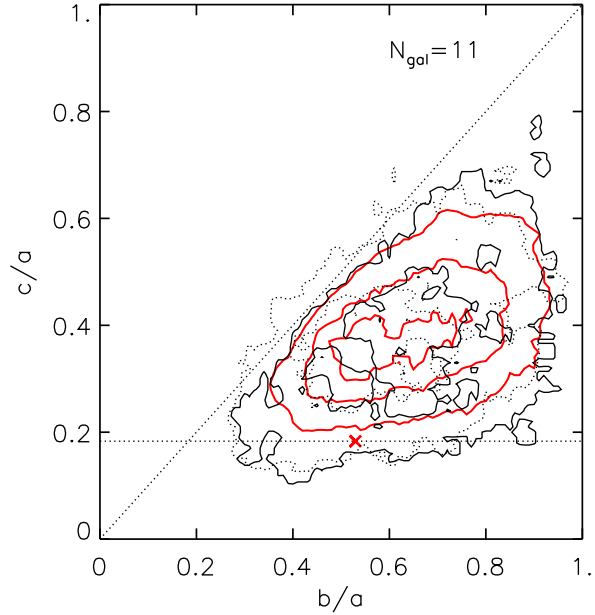}}\\%
\caption{The distribution of shape parameters for systems of 11
  satellites. The black lines show results for systems around 1686
  Milky Way like halos in the MRII and the dotted lines indicate the
  results after correcting for missing satellites in the
  galactic plane as described in the text. The red lines show results
  for 10000 randomly selected systems with the same radial
  distribution as the observed satellites of the Milky Way, but with a
  uniform distribution in solid angle. The three contour levels
  correspond to fractions of 30, 60 and 90 percent. The axis ratios of
  the 11 classical satellites of the Milky Way are indicated by the
  red cross, with the value of $c/a$ highlighted by the horizontal
  dotted line.}
\label{fig:Fig8}
\ec
\end{figure}

\begin{figure}
\bc
\hspace{1.4cm}
\resizebox{9cm}{!}{\includegraphics{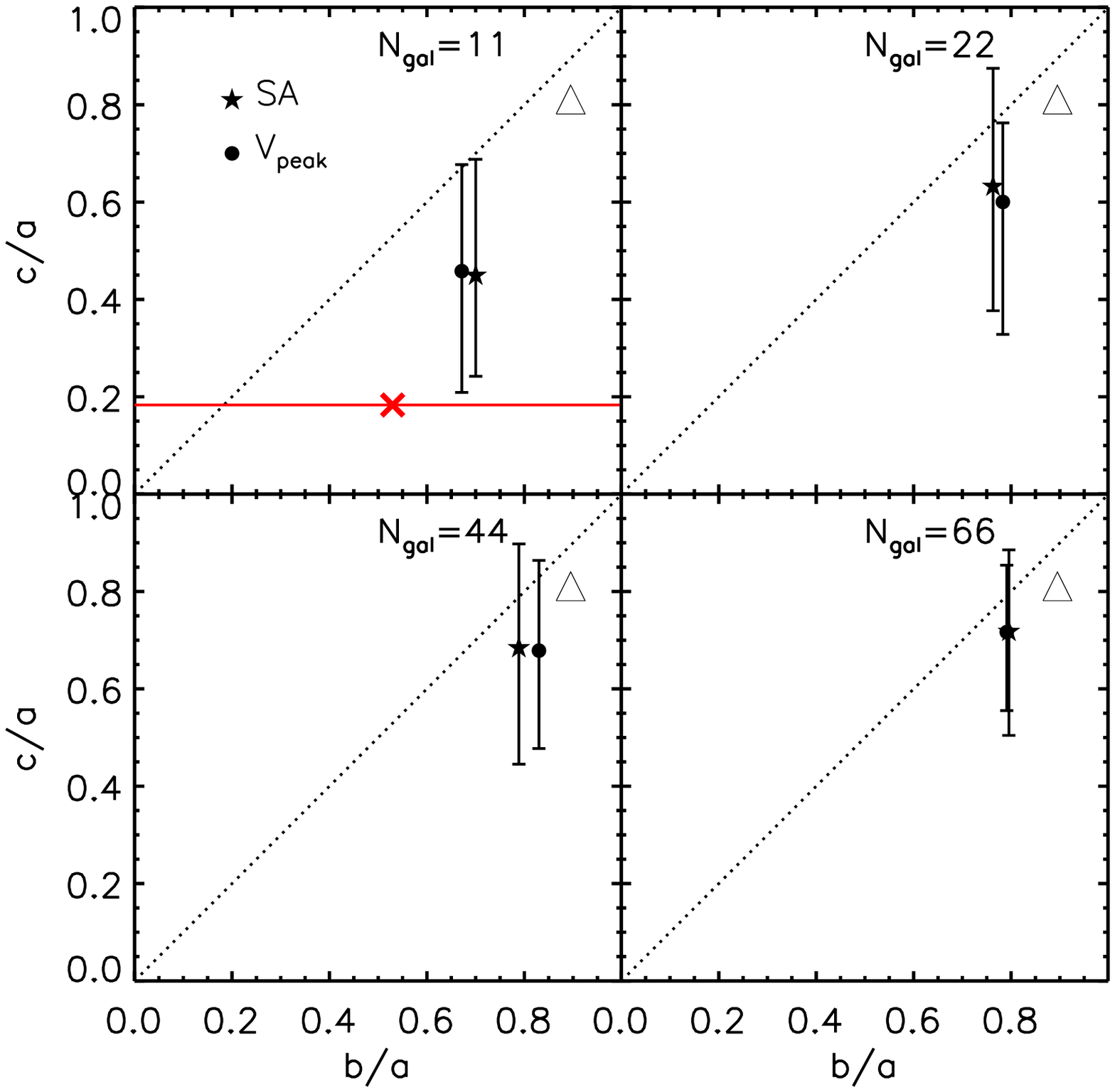}}\\%

\caption{Shape parameters for systems consisting of the top $N_{\rm gal}$
satellites according to our two models of satellite galaxy formation. The
symbols (asterisks for the semi-analytic model and filled circles for the $V_{\rm
peak}$ model) give the directly measured mean values, averaged over the 6
Aquarius halos. The error bars span the minimum and maximum values amongst the
6 halos after correction for satellites missing due to obscuration in
the Galactic plane. The four panels show results for $N_{\rm gal}=11, 22, 44$,
and 66. In the top-left panel, the values for the top 11 satellites of the
Milky Way ranked by luminosity are indicated by a red cross, with the value
of $c/a$ highlighted by a red horizontal solid line. 
  The triangle near the top-right corner of each panel shows the mean
  axis ratios of the entire dark matter distribution in the 6 Aquarius
  halos.}
\label{fig:Fig9}
\ec
\end{figure}

\subsection{The angular distribution of satellites and the great pancake}
\label{sec:pancake}

As we have seen in the preceding section, the semi-analytic and
$V_{\rm peak}$ models reproduce the observed radial profile of the 11
classical satellites of the Milky Way. In this subsection we
investigate if they also reproduce the highly flattened configuration
seen in the Milky Way, first noted by \cite{lyndenbell76} and
investigated more recently by
\citet{kroupa05,libeskind05,kang05,libeskind09} and \cite{deason11}. We use
similar techniques to those introduced by \cite{libeskind05} and
\cite{kang05} to describe the flattening of the satellite population,
namely, the overall shape of its distribution and its thickness in the
flattest dimension.

\subsubsection{The axis ratios of satellite systems} 
\label{sec:axisratios}

To determine the shape of a satellite system we calculate the moment
of inertia tensor of its members (weighting each of them equally) and
derive the principal axes of the distribution. The eigenvalues of the
diagonalised inertia tensor are proportional to the rms deviation of
the $x, y$ and $z$ coordinates of the system. Denoting the major,
intermediate and minor axes by $a$, $b$ and $c$ respectively ($a > b >
c$), the flattening of the system may be quantified by the axial
ratios $c/a$ and $b/a$. 

The axial ratios of the system consisting of the 11 classical
satellites of the Milky Way are indicated by a red cross in
Figs.~\ref{fig:Fig7} and~\ref{fig:Fig8}, with the value of $c/a=0.18$
highlighted by the horizontal dotted line.  This is slightly lower
than the value obtained by \cite{libeskind05} who had an error in
their computer code (N. Libeskind, private communication).  The
faintest of the 11 classical satellites are Draco and Ursa Minor
($M_V=-8.8)$, but Canes Venitici is just 0.2 mag fainter than this
\citep{McConnachie12}.  We have checked that including Canes Venitici
has virtually no effect in the estimated axis ratios of the system,
as shown by the red square in Fig.~\ref{fig:Fig7}.

The corrected value of $c/a$ for the Milky Way's
classical satellite appears quite extreme. However, with a sample of only 11
objects, it is not immediately obvious that a value even as low as $c/a=0.18$
is significant. To assess the significance of this result, we follow
\cite{libeskind05} and create a large number of artificial systems of 11
satellites in which the radial distances of the members are the same as the
distances of the real satellites, but the latitude and longitude are assigned
at random on the surface of a sphere. Isocontours of axial ratios for a set of
10,000 such isotropic systems, corresponding to number fractions of 30, 60 and
90 percent, are shown as red curves in Fig.~\ref{fig:Fig8}. We find that only
98 of these samples result in values of $c/a$ smaller than 0.18. Thus,
we conclude that the distribution of the 11 classical satellites of the Milky
Way is highly anisotropic, with only a 1 per cent probability that such a low
value of $c/a$ would result from a statistical fluctuation of an intrinsically
isotropic system.

Applying the same procedure to the simulations we obtain the axis ratios for
systems consisting of the top 11 satellites, ranked by stellar mass, that
lie within 250~kpc of the centre in our two models. The results are shown in
Fig.~\ref{fig:Fig7} for the six Aquarius halos, using asterisks for the
semi-analytic model and circles for the $V_{\rm peak}$ model. We also plot the
average axis ratios of the 6 Aquarius halos as a whole (within $r_{200}$) as a
large triangle.  The axis ratios of the 11 classical satellites of the Milky
Way are plotted as a red cross. Three conclusions emerge from this figure.
Firstly, as noted earlier by \cite{libeskind05} and \cite{kang05}, massive
satellite systems can be much flatter than the halo as a whole.  Secondly, the
two methods for ranking galaxies by stellar mass give very similar
results. Thirdly, the satellite systems in the simulations appear generally
less flattened than the 11 classical Milky Way satellites.

Although it is often assumed that the 11 classical satellites
represent a complete sample of bright satellites in the central
regions of the Milky Way, it is, of course, possible that bright
satellites might remain undetected in the zone of avoidance as a result
both of large extinction and confusion by foreground stars.  Assuming
that satellites are intrinsically uniformly distributed in $(1-{\rm
  sin}|b|)$, where $b$ denotes Galactic latitude, \citet{willman04}
estimated that the known population of dwarf satellites could be 33
percent incomplete because of obscuration. However, \citet{whiting07}
argued that, at most, there are one or two dwarfs still undiscovered
near the Galactic plane.

To assess the possible effects of obscuration, we assume a geometry for
the Galactic disk in our simulations and exclude satellites in the
corresponding zone of avoidance.  The ``great pancake'' in the Milky Way lies
perpendicular to the Galactic disk.  We therefore imagine that each of our
Aquarius halos has a ``Galactic plane'' whose normal is parallel to the major
axis of its satellite system. We then calculate the latitude of each satellite
relative to this plane. If the latitude is less than a critical value, $\pm
\theta_{\rm crit}$,  we regard this satellite as undetected and exclude it from
our sample of 11, replacing it by the next in the list. (For a 33 percent
occulted sky fraction, $\theta_{\rm crit}=9.5^{\circ}$.) We repeat this process
iteratively 2000 times, each time recalculating the shape of the current
satellite system and the position of the obscuring ``Galactic plane.'' In 2000
cases, satellites were hidden behind the model galactic plane and replaced, on
average, with 2.6 new satellites.

The results of this procedure, for $\theta_{\rm crit}=9.5^{\circ},$
are reflected in Fig.~\ref{fig:Fig7} as error bars spanning
the range of $c/a$, with the uncorrected sample shown as a filled
symbol. It is striking that the inferred shape of the satellite system
is extremely sensitive to the inclusion or exclusion of only one or
two members.  Allowing for sample incompleteness improves the
comparison between models and data noticeably, with several of the
models now coming close to the data.

To improve our statistics, we also analysed satellite systems in the MRII. As
discussed earlier, the $V_{\rm peak }$ model is strongly affected by the
relatively low resolution of the MRII, but not so the semi-analytic model,
provided satellites with dark matter haloes below the resolution limit 
are included in the analysis. As before, we consider all dark matter halos with
virial mass $1 \times 10^{12} \leq M_{\rm 200} \leq 2 \times 10^{12}$ that have
at least 11 satellites within $250$~kpc. This provides a sample of 1686
Milky Way halos. The axis ratios of the brightest 11 satellites in each of the
systems are shown as the black contours in Fig. ~\ref{fig:Fig8}, where levels
correspond to number fractions of 30, 60 and 90 percent. We find only 101
satellite systems with an axis ratio $c/a$ smaller than the observed value of
0.18, giving a probability of finding a configuration flatter than that in the
MW of only $101/1686 = 6\%$.

Just as for the Aquarius sample, we make a correction for the possible omission
of satellites near the Galactic disk. For each halo, we assume a Galactic plane
perpendicular to the minor axis of the satellite distribution, draw 2000 random
samples (replacing lost satellites and iterating) and obtain the mean values of
the axis ratios. As the dotted contour in Fig~\ref{fig:Fig8} shows, the
probability of finding a system in the simulations as flat or flatter than that
of the Milky Way is not sensitive to our corrections for obscuration by the
Galactic disk (under the assumption that the minor axes of the satellite system
and the disk are perpendicular).

The high mass resolution of the Aquarius simulations allows us to investigate
how the flattening of the satellite system varies as increasingly faint
satellites are included in the sample and thus to make predictions for the
larger samples that may become available in the future as surveys such as
Pan-STARRS1 and LSST discover new, fainter satellites in the Milky Way. The
axial ratios for systems of $N_{\rm gal}= 11, 22, 44$ and 66 satellites are
shown in the four panels of Fig.~\ref{fig:Fig9}.  The minimum and maximum
values amongst the 6 Aquarius halos, after correcting for the effects of the
zone of avoidance, as discussed above, are indicated by the error bars.
Fig.~\ref{fig:Fig9} shows that, as the size of the satellite sample increases,
the distribution becomes increasingly less flattened, in agreement with the
results of \citet{kang05}. When the sample size reaches 66, the shape of the
satellite configuration is very close to that of halo dark matter, indicated by
a large triangle in each panel.  This prediction from our simulations should be
readily testable with potentially forthcoming samples of new faint satellites,
provided their selection functions are well understood.

\subsubsection{The thickness of the satellite ``disk''} 
\label{sec:thickness}

\begin{figure}
\bc
\hspace{2.cm}
\resizebox{9cm}{!}{\includegraphics{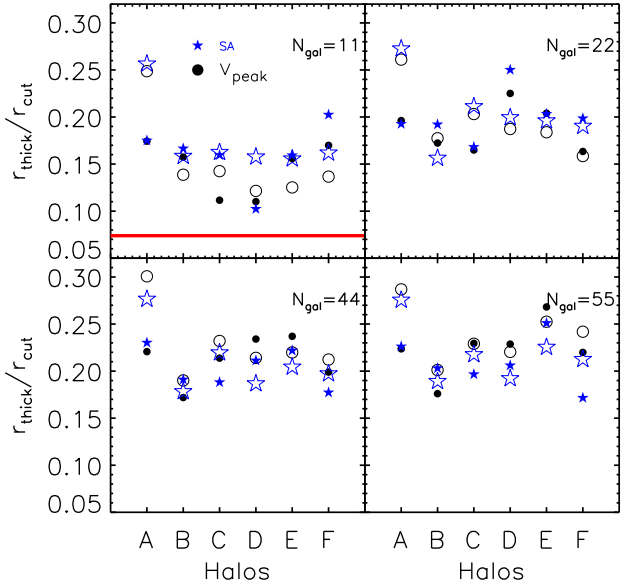}}\\%
\caption{The flattening (defined as the ratio of the {\it rms} height
  of the satellites relative to the best fit plane divided by the
  radial extent of the system) of the satellite distribution in the
  six Aquarius halos. Different panels correspond to samples
  containing the top $N_{\rm gal}$ satellites according to our two models of
  satellite formation, which are illustrated by the filled symbols, as
  indicated in the legend. The open symbols show the average of 20,000
  artificial samples each with the same radial distribution as the
  corresponding Aquarius sample but with random angular positions on
  the sky. In the $N_{\rm gal}=11$ panel the flattening of the 11 brightest
  satellites of the Milky Way, 0.074, is indicated by a red solid line.}
\label{fig:Fig10}
\ec
\end{figure}

\begin{figure}
\bc
\hspace{-1.4cm}
\resizebox{9cm}{!}{\includegraphics{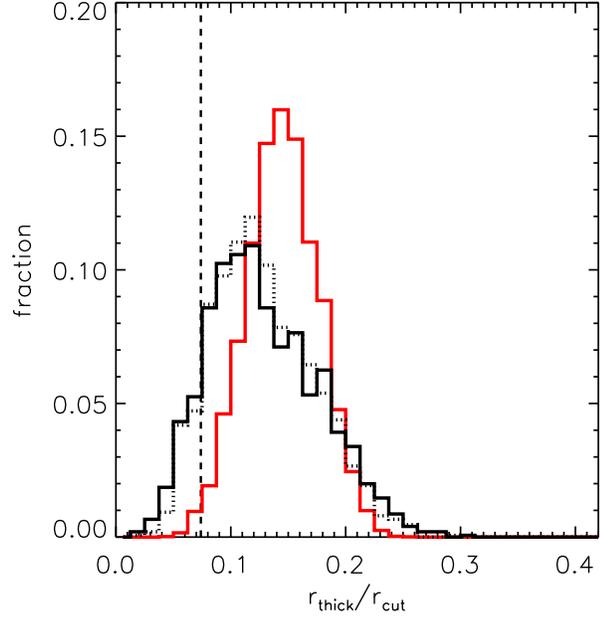}}\\%
\caption{The distribution of flattenings (obtained from the {\it rms}
height of satellites relative to the best-fit plane) for satellite
systems. The solid black line shows results for the 1686 halos in the MRII
with $1 \times 10^{12} \leq M_{\rm 200} \leq 2 \times 10^{12}$ that
have at least 11 satellites within $250$~kpc. The red histogram
corresponds to 10000 samples with the same radial distribution as the
11 classical satellites of the Milky Way and a random distribution of
angles on the sphere. The dotted histogram shows results for the MRII
corrected for the possible loss of satellites near the Galactic disk
taken from 2000 realizations.  
The flattening of the 11 classical satellites of
the Milky Way, 0.074, is shown as a dashed vertical line.}
\label{fig:Fig11} 
\ec
\end{figure}

An alternative method to estimate the flattening of the ``Great
pancake'' is to measure the thickness of the slab defined by the
pancake. Following \citet{kroupa05} and \citet{kang05}, we find the
best fit plane to a given satellite sample by minimizing the
root-mean-square of the height of each satellite relative to the
plane. The thickness of the slab is taken to be the {\it rms} height
about the best-fit plane and the ratio of the thickness to the radial
extent of the system, $R_{\rm cut}=250$~kpc, is used to characterise
its flattening. 

The flattening of the satellite disk in each of the six Aquarius halos
is shown in Fig.~\ref{fig:Fig10}. The four panels correspond to
systems with different numbers of members, $N_{\rm gal}$. Results for
our two different satellite galaxy formation models are indicated by
the filled symbols. The open symbols show results for 20,000
artificial samples with the same radial distribution as the
corresponding Aquarius sample but with random angular positions on the
sky.  In some cases (e.g. Aq-A in all panels, Aq-D in the $N_{\rm
  gal}=11$ panel, Aq-C in the $N_{\rm gal}=22$ panel), the simulated
satellite systems are significantly anisotropic but in others, the open
symbols lie close to the filled symbols, indicating that these systems
show no significant flattening according to this test.  In the top
left panel, we compare the model results with the measurement for the
11 classical satellites of the Milky Way, whose thickness (indicated
by the red line) is 0.074. This population is clearly flatter
than most of the simulations, although the satellite systems in Aq-C
and Aq-D come close.
The remaining panels in Fig.~\ref{fig:Fig10} show that, as was the
case for the axial ratios, systems containing more and more satellites
become progressively less flattened. 

We also estimated the thickness of satellite systems in the 1686 Milky
Way like halos found in the MRII as described in
Section~\ref{sec:axisratios}, which provides better statistics than the
Aquarius halos. The histogram of flattening values is shown in
Fig.~\ref{fig:Fig11}. The value measured for the 11
classical satellites of the Milky Way, 0.074, is indicated by the vertical
dashed line. This degree of flattening is highly significant: of 10000
artificial isotropic samples of 11 satellites having the same radial
distribution as the observed satellites (red histogram), only 107
are flatter than the Milky Way system. This corresponds to a probability of
only 1.07 percent. In the MRII a total of 219 cases have flatter
satellite systems than the Milky Way, corresponding to a probability
of 13 percent. When the effects of the Galactic disk are taken into
account, the distribution of flattenings is not significantly affected
(dotted histogram in Fig.~\ref{fig:Fig11}).

\begin{figure}
\bc \hspace{1.4cm}
\resizebox{9cm}{!}{\includegraphics{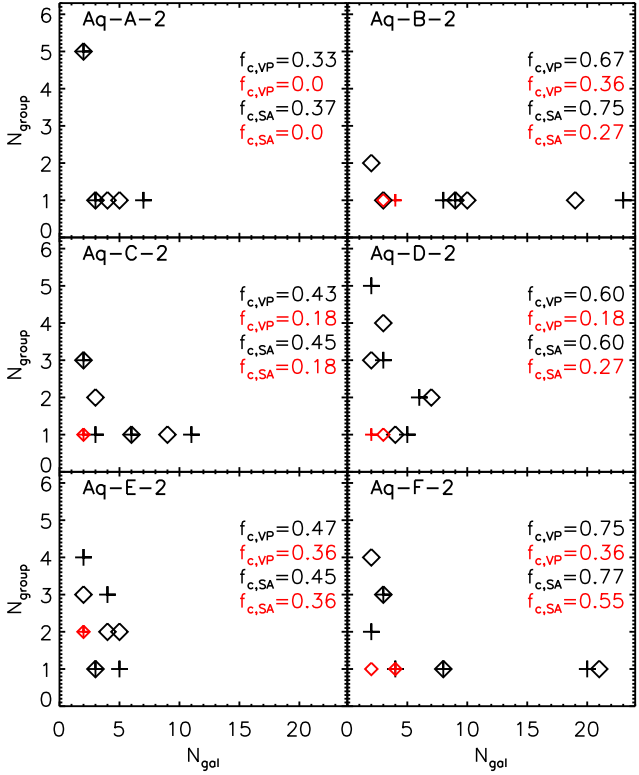}}\\%
\caption{Histogram of multiple accretion satellite events in the six
  Aquarius halos. Satellites identified by the semi-analytic model are
  shown as diamonds, while those identified by the $V_{\rm peak}$ model are
  shown as crosses. In both cases, red symbols correspond to the top
  11 satellites and black symbols to the top 60. The fractions of
  correlated halos relative to the total number of satellites,
  $f_{\rm c,SA} (N)$, for the semi-analytic model, and $f_{\rm c,VP}(N)$ for the
  $V_{\rm peak}$ model, where $N$ is the number of satellites in the
  system, are given in the legend.}
\label{fig:Fig12} 
\ec
\end{figure}

\begin{figure}
\bc \hspace{-1.4cm}
\resizebox{9cm}{!}{\includegraphics{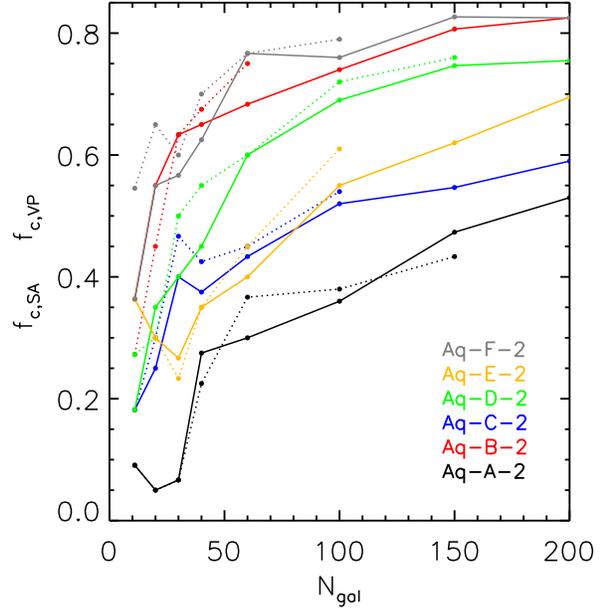}}\\%
\caption{The fraction of the top $N_{\rm gal}$ satellites which are
  accreted into the main halo in a group with at least one partner, as
  a function of $N_{\rm gal}$. The dotted lines show results for
  satellites in the semi-analytic model ($f_{c,SA}$), the solid lines
  for satellites in the $V_{\rm peak}$ model ($f_{c,VP}$).}
\label{fig:Fig13} \ec
\end{figure}

\subsection{The multiplicity of accreted groups}

As we have seen, the distribution of satellites in the Milky Way is
significantly anisotropic. \cite{libeskind05} ascribed this striking
property to the accretion of satellites along a few filaments of the
cosmic web.  However, in agreement with previous studies, we have
shown that, although flattened satellite systems are common, the Milky
Way's satellites are even more flattened than the expectation for
halos of similar mass. \cite{li08} noted that the natural flattening
induced by this anisotropic infall is strongest when considering
samples of galaxies accreted in `groups', defined by common infall
conditions (time, direction and orientation of orbital angular
momentum). Considering all dark matter subhalos regardless of their
mass, they found that samples of 11 subhalos drawn from only one such
infalling `group' have a configuration as flat as the classical Milky
Way satellite distribution in 73 percent of cases. This probability
falls to only 20 percent if only 5/11 satellites are drawn from a
single group. However, since Li \& Helmi studied only one
high-resolution halo, they were unable to quantify how likely it is
that the {\em most massive} 11 satellites of the Milky Way fell in in
just one or two groups and so to test whether or not this process is a
probable explanation for their unusually flat configuration. 

In order to explore the effect studied by \cite{li08} in our six
`typical' Milky Way halos, we identify satellites at the final time in the
simulation and trace them back to the friends-of-friends group to which they
belonged before they were accreted. If two or more satellites were part of the
same group at one or more subsequent simulation snapshots, we regard them as
correlated (\citealt{li08} did not require their correlated subhalos to be
members of the same FOF group, only to share similar infall times, directions
and orbital angular momentum orientations; in practice the two definitions are
likely to have similar results). In Fig.~\ref{fig:Fig12} we plot, for each
Aquarius halo, the number of groups in which a given number of satellites were
brought into the main halo (counting only satellites surviving to redshift
zero).  As before, we identify satellites using both our semi-analytic and
$V_{\rm peak}$ models (indicated by diamonds and crosses respectively) and we
consider samples of $N_{\rm gal}=11$ (red symbols) and $N_{\rm gal}=60$ (black
symbols) satellites.  The fractions of correlated halos relative to the total
number of satellites ($f_{\rm c,SA}$ for the semi-analytic model and $f_{\rm
c,VP}$ for the $V_{\rm peak}$ model) are given in the legend.

Focussing first on the top 11 satellites in each model (red symbols), we can see
that, except in the semi-analytic model for halo F, fewer than half of these
satellites are members of groups according to our definition; that is, more
than half are accreted singly into the main halo. This suggests that the
probability of the top 11 satellites falling into the halo as members of only
two or three groups is low; the moderate flattenings of the Aquarius satellite
systems represent the typical outcome of anisotropic accretion, and are not
enhanced by closer orbital associations between particular satellites.  Testing
whether or not the extreme flattening of the Milky Way's system can be
explained by the rare occurrence of a handful of `rich' or `top heavy'
accretion events, as proposed by \citet{li08}, will require a much larger suite
of simulations comparable to Aquarius. 

The prevalence of strongly correlated accretion increases as we consider larger
satellite samples. For example, in halos B and F more than 15 of the top 60
satellites were part of a single group before accretion, as may be seen in
Fig.~\ref{fig:Fig12}. The increase in $f_{\rm c,SA}$ and $f_{\rm c,VP}$
with the size of the satellite sample is illustrated in Fig.~\ref{fig:Fig13}
and is seen in both our methods for ranking satellites. For systems of only 10
satellites, the correlated fraction is $\sim 10-50 $ per cent, but this rises
to $\sim 30 - 80$ per cent for systems of 100 satellites. Our findings
reinforce the conclusion of \citet{lovell11} that a more complete census of
satellite motions in the Milky Way and M31 will reveal a significant fraction
with common orbital planes.

\section{Conclusion and Discussion}
\label{sec:conclusion}

The spatial distribution of satellites in the Milky Way and other
galaxies reflects both the nature of the dark matter and the processes
of galaxy formation. In this paper we have used high mass resolution
cosmological simulations of structure formation in the $\Lambda$CDM
model: the Aquarius resimulations of galactic scale dark matter halos
and the Millennium II simulation (MRII) of a cubic volume 100
$h^{-1}$Mpc on a side, with about 2000 halos of mass comparable to the
Milky Way's halo. We have further employed two models of galaxy
formation to trace satellites in the simulation: the semi-analytic
models of \cite{cooper10} and \cite{guo11} for Aquarius and MRII
respectively and a less sophisticated model in which galaxy stellar
mass is assumed to be proportional to $V_{\rm peak}$, the highest
circular velocity attained by the halo throughout its life. Our two
methods place 10 of the most massive 11 satellites in the same
Aquarius halos.

The combination of high-resolution simulations and robust galaxy formation
models allows us to identify satellites in a reliable way. We find that good
mass resolution is essential to obtain an accurate estimate of the radial
distribution of satellites. Indeed, even at the resolution of the MRII,
$m_p=6.9 \times 10^6 h^{-1} M_{\odot}$, many genuine substructures are
artificially destroyed, as may be seen in Fig.~\ref{fig:Fig5}. In this case,
the $V_{\rm peak}$ method for assigning stellar mass to satellites, as well as
all previous cosmological simulations which have poorer resolution
\citep[e.g.][]{libeskind05,kang05}, would give inaccurate results. By contrast,
our semi-analytic model, which continues to track satellites even after they
have lost their halos does give a faithful prediction of the theoretical
expectations in the $\Lambda$CDM model. In the case of the Aquarius
simulations, the resolution is good enough that galaxies with unresolved halos
galaxies are unimportant \citep{font11} and we can compare our
semi-analytic results with the $V_{\rm peak}$ method. The results from the two
ranking methods agree well.

Our main results concerning the satellite radial distribution are: (i)
there is substantial halo-to-halo scatter in the radial distribution
of satellites in the Aquarius simulations, but in all cases the 
massive satellites have a similar radial density profile to that of the dark
matter; (ii) both the semi-analytic and $V_{\rm peak}$ models give
results in good agreement with the measured radial distribution of the
11 classical satellites of the Milky Way; (iii) the radial density
profile of larger satellite samples, going down to lower masses, is
similar to that of the 11 most massive, at least up to samples of 100
satellites. This prediction may be tested by future surveys such as
Pan-STARRS. 

We also investigated the angular distribution of satellites. This is
interesting because the Milky Way's 11 brightest satellites show a
very anisotropic distribution, the ``Great pancake'' of
\cite{libeskind05}. We characterized the angular distribution by
measuring both the axial ratios of the satellite system and the
thickness of the slab in which the satellites are concentrated. We
verified that the peculiar distribution seen in the Milky Way is
highly significant: only 1 percent of isotropic samples with the same
radial distribution are flatter than the Milky Way's system. In the
1686 halos in the MRII which have a mass plausibly similar to that of
the Milky Way we find systems flatter than that of the classical
satellites with probabilities of 6 and 13 percent according to the
axial ratio and slab thickness tests respectively. These probabilities
are slightly lower than that found by \citet{libeskind05}, who used a
slightly incorrect value for the flattening of the Milky Way
satellites. Our probabilities are also lower than those found by by
\citet{kang05}, a discrepancy that is readily understood as a
consequence of the poor mass resolution of the simulations used by
these authors. The Milky Way system thus appears to be in the tail of
the flattening distribution expected for massive satellites in the
$\Lambda$CDM model. The presence of a satellite as massive as the LMC
\citep{boylan10,busha11,guoq11,liu11} and the polar alignment
of the system \citep{deason11} seem to be comparably rare, but still
consistent with the predicted distributions.

Finally, we investigated the extent to which satellites are accreted in groups.
This is interesting for several reasons, including the possibility that the
Great pancake might be explained by multiple accretion in 2-3 groups
\citep{li08}. Our simulations confirm that this is a rare occurrence in
$\Lambda$CDM. On average, only 30 percent
of the top 11 satellites in the Aquarius simulations share a
friends-of-friends group with another top 11 satellite before infall; the
rest come into the main halo without any companions of comparable mass.
However, multiple accretion becomes increasingly important for larger, fainter
samples of satellites. For example, in samples of the 60 most massive
satellites in two of the Aquarius halos (Aq-B and Aq-F), we find that as many
as 20 come into the main halo in a single group and as many as 11 in the other
simulated halos. This interesting property may potentially have observational
consequences \citep[e.g.][]{sales12}.

\cite{libeskind05} proposed that filamentary accretion is responsible
for the flattening of Milky Way-like satellite populations.
\cite{lovell11} subsequently showed that the subhalos of the Aquarius
simulations have strongly correlated orbital angular momenta as the
result of anisotropic accretion.  Recently, however,
\citet{pawlowski12} re-analysed the angular momentum orientations of
subhalos in Aquarius (as presented in \citealt{lovell11}) and
concluded that anisotropic accretion is unimportant for producing
flattened satellite systems, in direct contradiction with the results
of \cite{libeskind05} and \cite{lovell11}.  They required that the
Milky Way flattening should be reproduced in the mean, rather than
being merely consistent with the expected distribution. As the Milky
Way is in the $\sim10$ per cent tail of the distribution predicted by
$\Lambda$CDM (as our study and others have shown), they reject a CDM
model in favour of tidal galaxy formation, which they claim readily
produces highly correlated disks of satellites. We believe that the explanation
offered by \cite{libeskind05} remains the most appropriate. A single
randomly chosen system such as the Milky Way is extremely unlikely
perfectly to represent the mean value of every measurable property. A
larger sample of satellites around other galaxies will test the tidal
formation hypothesis of \citet{pawlowski12} in which highly flattened
configurations are easily achieved and should therefore be the norm.
If, on the other hand, the CDM model is a realistic description of
nature, then the average satellite configurations should be only
moderately flattened, as illustrated in Figs.~\ref{fig:Fig7}
and~\ref{fig:Fig8}.

Our simulations are useful not only to test our models against data
for the classical satellites of the Milky Way, as we have done here,
but also to make predictions for future surveys that will quantify the
spatial distribution of larger and fainter satellite samples, both in
the Milky Way and in other galaxies.  For a sample of Milky Way
satellites complete to very faint magnitudes, we have shown that the
luminosity function \citep{font11}, radial distribution and system
shape should vary less from halo to halo than is the case for the most
massive eleven. Measuring these three highly characteristic properties
of the satellite system test both the $\Lambda$CDM cosmology and
models of galaxy formation in novel, interesting ways.

%%%%%%%%%%%%%%%%%%%%%%%%%%%%%%%%%%%%%%%%%%%%%%%%%%%%%%%%%%%%%%%%%%%
\section*{Acknowledgements}

We thank Julio Navarro and Simon White for useful suggestions and
comments, and Qi Guo for 
useful discussions and explaination of the MRII galaxy catalogue. The
simulations of the Aquarius Project were carried out at the Leibniz Computing
Center, Garching, Germany, at the Computing Centre of the Max-Planck-Society in
Garching, at the Institute for Computational Cosmology in Durham, and on the
`STELLA' supercomputer of the LOFAR experiment at the University of Groningen.
JW acknowledges a Royal Society Newton International Fellowship, CSF a Royal
Society Wolfson Research Merit Award and ERC Advanced Investigator grant,
COSMIWAY. This work was supported by an STFC rolling grant to the Institute for
Computational Cosmology.

%%%%%%%%%%%%%%%%%%%%%%%%%%%%%%%%%%%%%%%%%%%%%%%%%%%%%%%%%%%%%%%%%%%%
\bsp
\label{lastpage}

\bibliographystyle{mn2e}
\bibliography{sat_v5}

\end{document}